\documentclass[twocolumn,showpacs,preprintnumbers,amsmath,amssymb,prb]{revtex4}
\usepackage{graphicx}
\usepackage{dcolumn} 
\usepackage{bm} 
\usepackage[hypertex]{hyperref}
\usepackage{longtable}

\begin{document}

\title{Controlling the effective mass of quantum well states in Pb/Si(111) by interface engineering}
\author{Bartosz Slomski$^{1,2}$}
\author{Fabian Meier$^{1,2}$}
\author{J\"urg Osterwalder$^{1}$}
\author{J. Hugo Dil$^{1,2}$}
\affiliation{
$^{1}$Physik-Institut, Universit\"at Z\"urich, Winterthurerstrasse 190, 
CH-8057 Z\"urich, Switzerland 
\\ 
$^{2}$ Swiss Light Source, Paul Scherrer Institut, CH-5232 Villigen, 
Switzerland}
\date{\today}

\begin{abstract}
The in-plane effective mass of quantum well states in thin Pb films on a Bi reconstructed Si(111) surface is studied by angle-resolved photoemission spectroscopy. It is found that this effective mass is a factor of three lower than the unusually high values reported for Pb films grown on a Pb reconstructed Si(111) surface. Through a quantitative low-energy electron diffraction analysis the change in effective mass as a function of coverage and for the different interfaces is linked to a change of around 2 \% in the in-plane lattice constant. To corroborate this correlation, density functional theory calculations were performed on freestanding Pb slabs with different in-plane lattice constants. These calculations show an anomalous dependence of the effective mass on the lattice constant including a change of sign for values close to the lattice constant of Si(111). This unexpected relation is due to a combination of reduced orbital overlap of the $6p_z$ states and altered hybridization between the $6p_z$ and $6p_{xy}$ derived quantum well states. Furthermore it is shown by core level spectroscopy that the Pb films are structurally and temporally stable at temperatures below 100 K.
\end{abstract}

\pacs{73.20.At, 73.21.Fg, 79.70.Dp}

\maketitle
\section*{INTRODUCTION}
The electronic structure of crystalline thin metal films is often dominated by the formation of standing electron waves in the direction perpendicular to the surface. The basic physics behind these quantum well states (QWS) is well understood \cite{Chiang:2000} and it has been found that a variety of properties follow the predicted dependence on the exact thickness of these films \cite{Schulte:1976}. The growth conditions can depend significantly on the substrate crystal orientation \cite{Dil:2010JPCM} and on the interface reconstruction \cite{Yeh:2000}. Furthermore the energies of the confined states can be altered via changing the atomic species at the interface \cite{Ricci:2004}. In this work we will extend this research and show that also the band dispersion of quantum well states can be altered by interface engineering, exemplified by the system of ultra-thin Pb films on Si(111) .

The structural and electronic properties of Pb on Si(111) have been studied for almost 30 years. The initial interest was focussed on the contact between a superconductor (Pb) and a semiconductor (Si). However, the rich variety of unexpected phenomena which were observed has incited research also in temperature ranges far above the superconduction transition temperature $T_{C}$ of Pb. Examples of these observations are the strong dependence of the Schottky barrier on the exact atomic structure \cite{Heslinga:1990}, the formation of magic height islands \cite{Budde:2000}, anomalies in the superconductivity transition temperature \cite{Guo:2004,Brun:2009,Eom:2006}, oscillations of the work function and extremely long excited state life times \cite{Kirchmann:2007,Kirchmann:2008}, and the observation of a Rashba-type spin splitting \cite{Dil:2008}. The origin of many of these phenomena lies in the peculiar electronic structure of Pb on Si(111) manifested in the high in-plane effective mass \cite{Mans:2002,Upton:2005,Dil:2006}. A control of this effective mass can thus also give control over many of the other physical properties which make Pb films on Si such a rich research subject.

\section{EXPERIMENT}
\begin{figure}[!h]
\begin{center}
\includegraphics[width=0.45\textwidth]{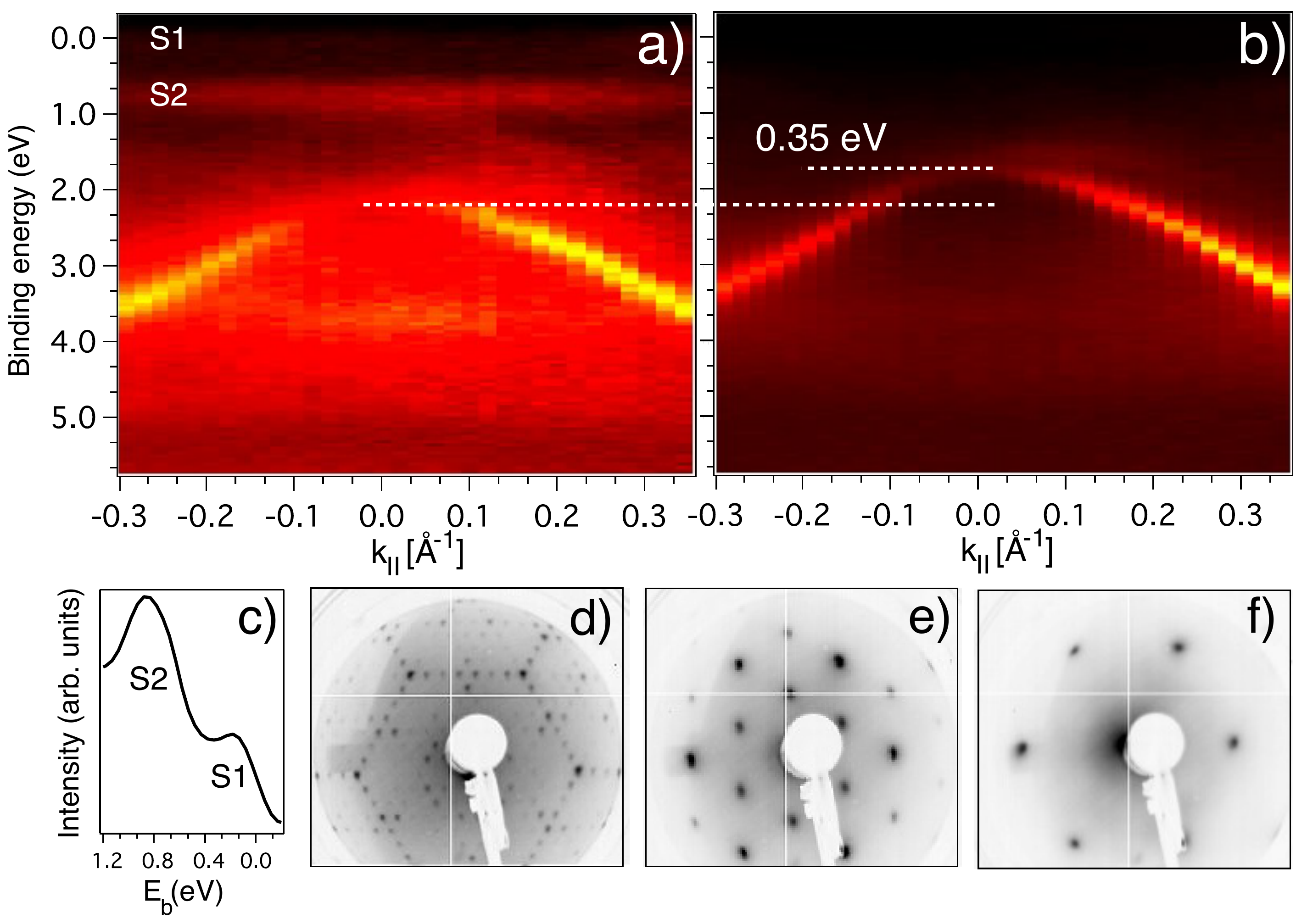}
\caption{(Color online) Band dispersion of (a) \emph{n}-Si(111)-(7$\times$7) and (b) \emph{n}-Si(111)-$(\sqrt{3}\times \sqrt{3})$-Bi$(\beta)$ along the $\overline{\Gamma \text{M}}$ direction. (c) Energy spectrum of \emph{n}-Si(111)-(7$\times$7) at normal emission. (d), (e) and (f) LEED patterns from the  (7$\times$7), Si(111)-$(\sqrt{3}\times \sqrt{3})$-Bi$(\beta)$ and from the (1$\times$1) surface of  Pb/Bi/Si(111).} 
\label{EDC_Si7by7}
\end{center}
\end{figure}

The angle-resolved photoemission spectroscopy (ARPES) and low energy electron diffraction (LEED) experiments were performed with the COPHEE setup at the Swiss Light Source of the Paul-Scherrer-Institut at the Surface and Interface Beamline. The energy and angular resolution of the spectrometer were set to 30 meV and 0.5$^{\circ}$, respectively \cite{Hoesch:Thesis}. In the current experiment the spin-resolution was not used. All valence band photoemission data were recorded at a photon energy of 24 eV and the core-level data at 70 eV using horizontally polarized light.
The base pressure of the chamber was below 2$\times$10$^{-10}$ mbar and the sample temperature during the measurements was T $<$ 70 K. The $n$-type Si(111) (44-62 $\Omega$cm) sample was degassed at 600 K for 24 h and flashed several times above 1300 K to remove adsorbates and to form the (7$\times$7) surface reconstruction. The cleanliness of the sample was checked with ARPES, where the surface states $S1$ and $S2$ at binding energies $E_{B} = 0.20$ eV and 0.82 eV could be identified, as shown in Fig.~\ref{EDC_Si7by7}(a) and (c). The LEED pattern displayed in Fig.~\ref{EDC_Si7by7}(d) confirms the (7$\times$7) surface reconstruction showing intense superstructure spots.

Bi and Pb interfaces were prepared through the deposition of approximately 3 monolayers (ML) of Bi or Pb from a water-cooled Knudsen cell onto the clean Si(111)-(7$\times$7) surface at low temperature and subsequent annealing until the $(\sqrt{3}\times \sqrt{3})$R30$^{\circ}$ (henceforth $\sqrt{3}$) surface reconstruction appeared in LEED as shown in Fig.~\ref{EDC_Si7by7}(e). From previous studies it is known that Bi has two $\sqrt{3}$ superlattice phases on Si(111) which form depending on the initial coverage and the annealing process \cite{Wan:91PRB}. The $\alpha$-phase forms at T $>$ 640 K with $\frac{1}{3}$ ML coverage in substrate units (1 ML = 7.83 $\times$ 10$^{14}$ atoms/cm$^{2}$) whereas the denser $\beta$-phase forms for temperatures below 640 K and has a coverage of 1 ML. In our experiment the annealing temperature was around 600 K, hence we conclude that the $\beta$-phase is formed. After creation of this phase the surface states disappear and the top of the valence band (VB) of Si is shifted by 0.35 eV to lower binding energy from initially 2.1 eV to 1.75 eV, which is apparent from a comparison of Figs.~\ref{EDC_Si7by7}(a) and (b). This band shifting indicates that a dipole layer is created.

The Pb $\sqrt{3}$ interface was formed under similar conditions, yielding the $\frac{4}{3}$ ML $\alpha$-phase, on which high quality Pb films can be grown \cite{Hupalo:2001}. It should be noted that the nomenclature is inverted and that the Bi $\beta$-phase and the Pb $\alpha$-phase are very similar in the sense that both form a trimer-type structure. To avoid confusion we will refer to both phases as the dense phase, based on the higher coverage compared to the competing phases.

On these respective $\sqrt{3}$ substrates thin crystalline Pb films were grown by deposition at a rate of $\frac{1}{3}$ ML per minute after cooling down again to below 100 K. For the Pb $\sqrt{3}$-phase it is known that, under these growth conditions, high quality Pb films can be grown with monolayer resolution which support quantum well states \cite{Yeh:2000,Upton:2004,Dil:2006}. One of the findings in this work is that also Pb layers deposited on the Bi interface accommodate well defined quantum well states. 

\section{The role of the metal to substrate interface}
In a first approximation the confinement of the conduction electrons within the metal film can be treated similarly to the well-known one-dimensional particle-in-a-box model. In this model the confinement of a particle by infinite potential barriers leads to discrete energy levels with standing wave solutions having nodes exactly at the boundaries of the box. A more realistic treatment of this problem is to approximate the boundaries by a finite barrier height. In this case the wave function describing the particle will have a certain penetration depth into the barrier specified by a phase shift \cite{Chiang:2000}. The main difference between both approximations is that in the case of a finite barrier the penetration of the wave function will lead to a larger effective width $d$ of the confinement box and therefore to a lower energy of the particle since $E \propto d^{-2}$ (for the same quantum number).

The description of QWS as standing waves allows to apply the so-called phase accumulation model in order to calculate the energies of such electronic states \cite{Smith:1985}.
A standing wave may form when the total accumulated phase $\Phi_{T}$ is an integer number $n$ of 2$\pi$. Here the total phase of the wave can be split into a contribution from the propagation through the overlayer $2d k_{z}$, when the electron travels back and forth through the medium, and a phase shift upon reflection at the metal to vacuum interface $\Phi_{V}$ and at the metal to substrate interface $\Phi_{S}$ \cite{Echenique:1978}:
\begin{eqnarray}
\Phi_{T} = \Phi_{V} + \Phi_{S} + 2d_{0}Nk_{z} \stackrel{!}{=} 2\pi n
\end{eqnarray} 
where $d_{0}$ is the intralayer spacing (Pb, $d_{0}=2.85$ \AA) and $N$ denotes the number of monolayers (the thickness of the overlayer is an integral number of $d_{0}$). In this approach the potentials at the metal to substrate and metal to vacuum interface are simply modeled as a phase shift. The exact nature of the potentials need not necessarily to be known.

While the metal to vacuum interface produces always the same phase shift, the following examples demonstrate the influences of the metal to substrate interface on QWS properties. 
Ricci \emph{et al.} \cite{Ricci:79B} studied QWS in thin Pb films on three differently terminated Si(111) surfaces, namely In, Au and Pb. 
Although In and Pb induce similar $\sqrt{3}$ reconstructions on Si(111), it was shown that the binding energies differ by $\approx$ 1 eV among the two interfaces for the same quantum number $n$ and thickness $d_{0}N$. A simultaneous fit of the measured energies using the phase accumulation model together with the interface phase shift according to $\Phi_{S}(E) = A + B\sqrt{E-E_{0}}\Theta(E-E_{0})$ allows to quantify the influence of the interfaces. Here $E_{0}$ is the energy of the valence band maximum, $\Theta$ the Heavyside function and $B$ a constant related to the electronic structure of Si. For Pb the fit yielded $A = 2.21$ and for In $A = -1.70$. The phase shift difference of both interfaces $|\Phi_{S}^{Pb}(E) - \Phi_{S}^{In}(E)| \approx \pi$, suggesting that the Pb interface induced phase shift $\Phi_{S}^{Pb}$ is over the entire confinement energy region larger than the phase shift induced by the In interface. As a result QWS on the Pb interface are lower in energy, because the effective width of the confinement box is larger. In a similar experiment \cite{Ricci:2004} the influence of the Schottky barrier on the confinement of QWS was investigated. QWS with energies $E$ within the band gap of Si(111) cannot couple to any bulk states thus an electron impinging on the interface will be perfectly reflected. The interface reflectivity is close to one, resulting in narrow photoemission line shapes. On the other hand electronic states with energies outside the gap, commonly referred to as quantum resonances (QR), give rise to broad peaks due to resonant coupling. By measuring the line width with ARPES as a function of QWS binding energy it is possible to determine the confinement edge $E_{0}$ and thus the transition from QWS to QR. For the Pb interface electronic states with $E_{B} < 0.55$ eV are truly confined whereas for the In interface sharp peaks can be observed down to $E_{B} < 0.58$ eV. These findings are consistent with the determined Schottky barrier, which is 0.62 eV for the Pb and 0.55 eV for the In interface. 

To summarize, interfaces of different types of atoms have a significant influence on the phase shift of the wave function at the metal-to-substrate interface and thus on the electronic properties of QWS. In this work, we will show that the interface has also an influence on structural properties of the metallic overlayer and consequently on the electronic structure.

\section{The in-plane dispersion of QWS}
The unusual flat in-plane dispersion of Pb QWS on Pb-$\sqrt{3}$/Si(111) is still a subject of discussion. Before we present our results and discussion we will first review some previous measurements of band dispersions of Pb QWS on different types of interfaces.

ARPES is a suitable technique to investigate the in-plane dispersion $E(k_{\parallel} )$ of QWS. The confinement of the states is present in the direction perpendicular to the surface whereas within the film plane the crystal can be regarded as infinite. Therefore the band structure around the surface Brillouin zone center (SBZC) $\bar{\Gamma}$~($\vec{k}_{||}$~=~0) is well described by a free electron gas according to $E(k_{\parallel} )~=~\frac{\hbar^2}{2m^{\star}}k^2_{\parallel}$ with $m^{\star}$ as effective mass and $\textbf{k}_{\parallel}=(k_{x}, k_{y}$) as the parallel momentum.

It is known that the dispersion of a band can be attributed to the degree of localization of the state; localized electronic states have less dispersion than delocalized states. The degree of localization is in simple systems related to the degree of overlap between the orbitals of which the state is composed: a small orbital overlap results in a high degree of localization and vice versa.
  
An example of a system with almost no interaction between the substrate and the overlayer is given by Pb on graphitized SiC \cite{Dil:2007}. The dispersion of the QWS is free electron-like with an effective mass of approximately $m_{e}$ (free electron mass) around the SBZC even for a low film thickness of 2 ML, where typically strain due to the lattice mismatch with the substrate is expected. Density functional theory (DFT) slab calculations are able to reproduce the measured dispersion by assuming a free-standing Pb slab. In this system the inert graphene layer decouples the influence of the substrate. On the other hand Pb deposited on Pb-$\sqrt{3}$/Si(111) (henceforth Pb/Pb/Si(111)) is an example of a strong influence of the substrate \cite{Upton:2004,Dil:2006}. For coverages up to 25 ML the states around SBZC have unusually high effective masses of up to $m^{\star}$ $\approx$ 10 $m_{e}$, which can not be modeled with a free-standing film assuming a bulk Pb in-plane lattice constant.  Here we will show that the Bi interface acts similarly to the graphene layer to decouple the Pb overlayer from the substrate and that the dispersion increases. It will be argued that the strong in-plane localization observed in Pb layers on the Pb-$\sqrt{3}$/Si(111) substrate can be explained by an increased atomic spacing in the direction parallel to the surface induced by the substrate.

\section{RESULTS AND DISCUSSION}
Before we discuss the properties of the QWS in the Pb films on the Bi-$\sqrt{3}$ interface we will first show that this interface is stable against intermixing with the Pb overlayer and that the properties of the QWS do not change significantly on the time scale of days.

\begin{figure}[!h]
\begin{center}
\includegraphics[width=0.45\textwidth]{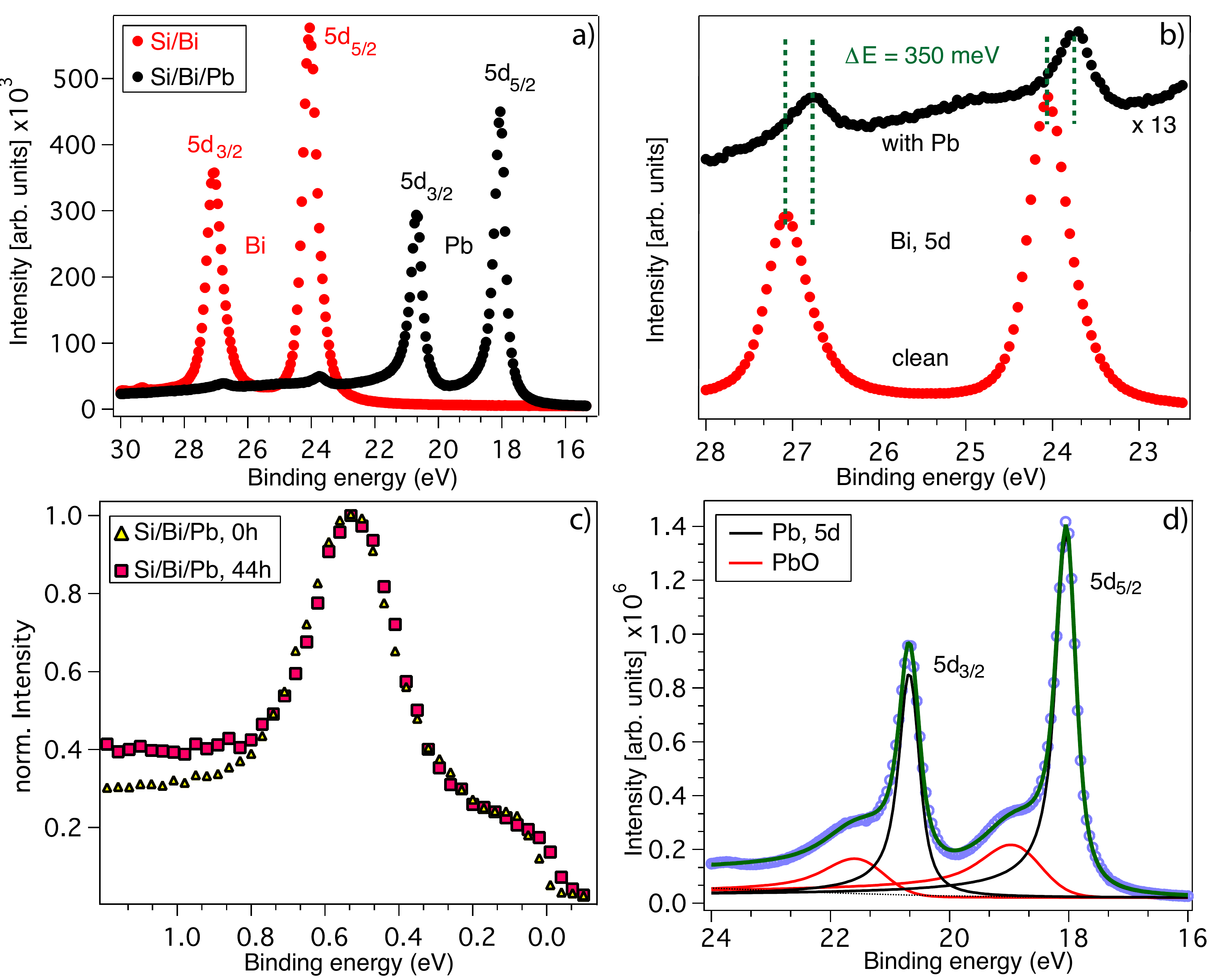}
\caption{(Color online) (a) Photoemission of Pb and Bi 5d core-levels from Bi-$\sqrt{3}$/Si(111) (black) and of 9 ML Pb/Bi/Si(111) (red/grey). Both heavy metals $(Z_{Pb}$ = 82, $Z_{Bi}$ = 83) show strong spin-orbit split  d-states. (b) Core-levels of Bi before (red/grey dots) and after Pb deposition (black dots). The Pb overlayer induces a chemical shift by 350 meV towards lower binding energy. The curves are shifted on the vertical scale for clarity. (c) Spectrum of a QWS shortly after preparation (triangles) and after 44 h (squares). (d) 5d Core-levels of Pb 24 h after preparation fitted with the clean and oxidized Pb components.}
\label{Corelevel}
\end{center}
\end{figure}

Each Bi atom of the interface binds covalently with one Si atom of the substrate and with two neighboring Bi atoms forming the so-called trimer centered at the $T_{4}$ site \cite{Cheng:1997}. The bonding to the Si atom is stronger than the bondings within the trimer, because with increasing substrate temperature the trimer-phase can be transferred into the $\alpha$-phase, where only one Bi atom is bound to three Si surface atoms. To investigate possible intermixing we used X-ray Photoelectron Spectroscopy (XPS), see Figure \ref{Corelevel}(a), to map the $5d$ core-levels of Bi and Pb. 

The red (grey) spectrum corresponds to  Bi-$\sqrt{3}$/Si(111) and the black spectrum to 9 ML of Pb on this substrate. The 5$d_{3/2}$ and 5$d_{5/2}$ core-levels of Pb and Bi are easily identified and labeled accordingly in Fig. \ref{Corelevel}(a). The intensities of the Bi core-levels are drastically reduced after Pb deposition, which is a first indication that no intermixing occurs at the interface. The remaining Bi intensity is either due to the fact that the Pb layer is not fully closed and small regions with the bare substrate remain visible, or due to electrons emitted from Bi atoms at the interface. To quantify what Pb thickness one would require for the second scenario the spectrum after Pb deposition was further analysed according to Ref.~\cite{Ertl:1974} with the Bi interface taken as an 1 ML substrate and the Pb overlayer as a thin film:
\begin{eqnarray}
\frac{I_{\text{Pb}}\sigma_{\text{Bi}}N_{\text{Bi}}^{0}}{I_{\text{Bi}}\sigma_{\text{Pb}}N_{\text{Pb}}^{0}} = \frac{1-\exp{\frac{-d_{\text{Pb}}}{\lambda_{\text{Pb}}}}}{\exp{\frac{-d_{\text{Pb}}}{\lambda_{\text{Pb}}}}-\exp{\frac{-d_{\text{Bi}}}{\lambda_{\text{Pb}}}}}.
\label{core}
\end{eqnarray}
Here $I_{\text{Pb}}$ and $I_{\text{Bi}}$ are the measured intensities, $\sigma$ is the photoionization cross section \cite{Crosssection:1993}, for Pb $\sigma_{\text{Pb}}$ = 21.55 Mbarn and for Bi $\sigma_{\text{Bi}}$ = 28.03 Mbarn, $N$ is the surface atomic density with $N_{\text{Pb}}^{0}$~=~9.43~$\cdot 10^{14}$ atoms/cm$^{2}$ and $N_{\text{Bi}}^{0}$ = 7.83 $\cdot 10^{14}$ atoms/cm$^{2}$. $d_{\text{Pb}}$ and $d_{\text{Bi}}$ are the thicknesses of the Pb overlayers and of the Bi layer. $\lambda_{\text{Pb}}$ = 5 \AA \  is the mean free path of the photoelectrons in the Pb layers. From Eq.~\ref{core} we obtain a thickness $d_{\text{Pb}}$ = 18 \AA \ which corresponds to 6.1 ML of Pb. From the analysis of the binding energy of the QWS, which provides an intrinsic thickness calibration, we obtain a thickness of 9 ML. In case of intermixing the ratio of the $5d$ core-level signals of Pb and Bi would be  I$_{\text{Pb}}$:I$_{\text{Bi}}$ = 6:1 for the same amount of Bi in a 9 ML Pb film, and thus significantly smaller than the measured ratio of 56 : 1. We therefore conclude that the Bi interface is stable against intermixing with Pb at the low temperatures under investigation. This result is not suprising since the bonding of the Bi to the Si substrate lowers the surface energy by filling dangling bonds. Although the Bi does not intermix with the Pb, its chemical environment is altered due to the Pb layer on top. As shown in Fig. \ref{Corelevel}(b) this results in a shift of the Bi core-levels by about 350 meV to lower binding energies.

A further advantage of the investigated ultra-thin films is their stability over time. Figure \ref{Corelevel}(c) shows energy distribution curves (EDCs) of a QWS shortly after preparation (triangles) and after 44 h (squares) exposure to the residual gas of the ultra-high vacuum chamber. The $5d$ core-levels of Pb (see Figure \ref{Corelevel}(d)) show an additional shoulder after 24 hours of measurement,  which can be attributed to oxidation of the Pb top layer. However, the peak width and the binding energy of the QWS are not affected by this oxidation and remain constant over time. The only quantity that changes is the peak-to-background ratio, which decreases from $1:0.3$ to $1:0.4$  presumably due to an increased number of adsorbates on the surface. This temporal stability of the system is advantageous as it allows to perform long time measurements, but it also indicates that the QWS survive a significant contamination as long as the sample is kept at low temperature. One would expect a change of the metal-to-vacuum phase shift, when the topmost layer becomes oxidized which in turn should influence the binding energy of the QWS. However, Peng \emph{et al.} have studied temperature dependent oxidation of the Pb(111) surface with STM \cite{Peng:2009} and found that at low temperature ($\approx$ 80 K) and low coverage ($<$~6~L) only small-size oxide clusters form. These clusters will serve as point defects resulting in an increase of the background without significant change in the confinement of the QWS. The oxidation processes may have an influence on $\Phi_{V}$ at higher coverages when a closed oxide layer forms \cite{dftox:2010}.

\begin{figure}[!h]
\begin{center}
\includegraphics[width=0.5\textwidth]{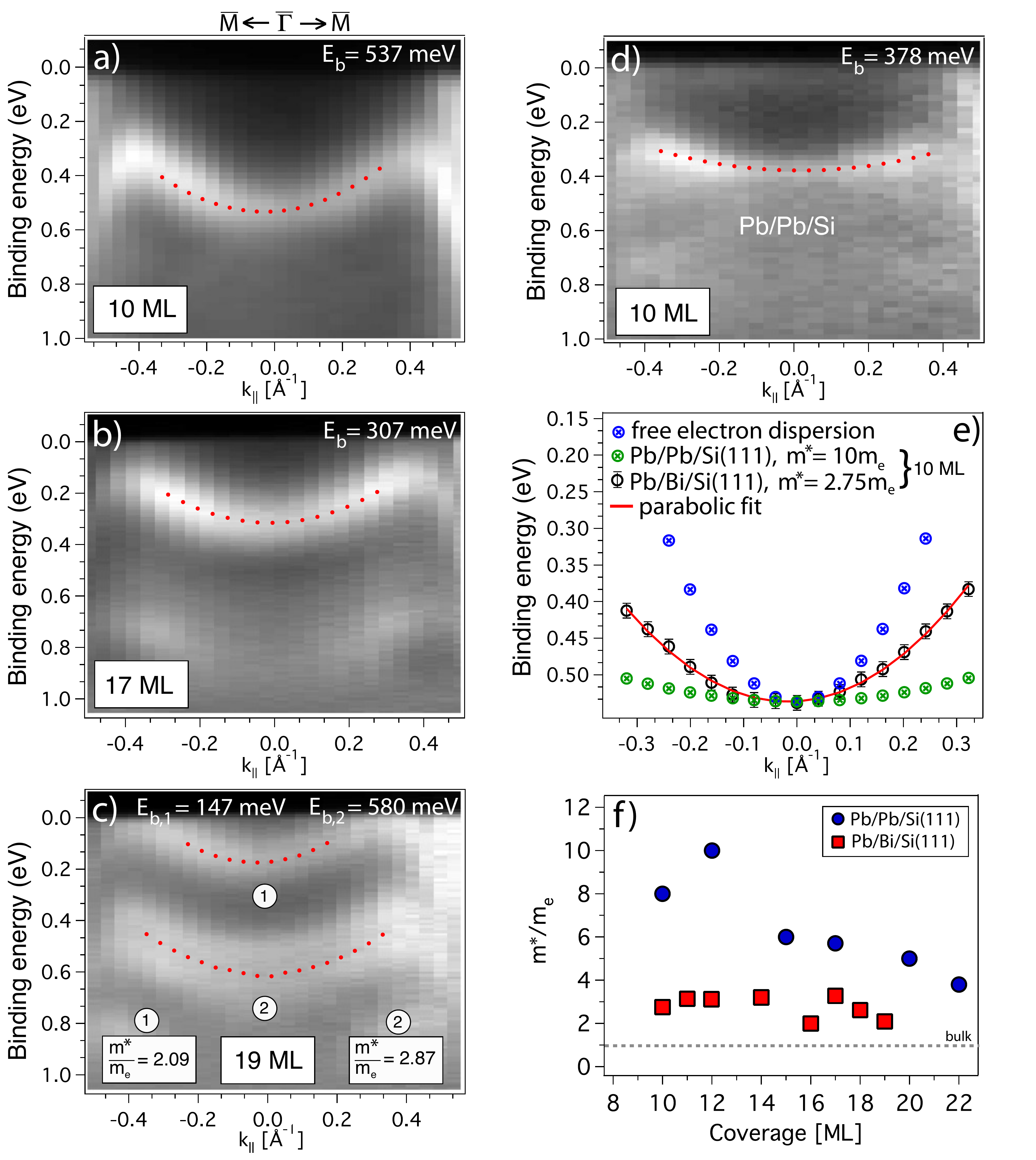}
\caption{(Color online) (a-c) Band dispersion of QWS in 10, 17 and 19 ML Pb films on Bi/Si(111). (d) Band dispersion of Pb QWS on Pb/Si(111). (e) Extracted dispersion (red dots) from Fig. \ref{QWS}(a) as described in the text. For comparison a free electron-like dispersion and the dispersion of Pb/Pb/Si(111) are included. (f) Red squares: Effective masses of QWS for different thicknesses of Pb deposited on the Bi/Si(111) obtained from the fitting procedure as described in the text. Blue circles: effective masses of Pb on Pb/Si(111) taken from \cite{Dil:2006}.}
\label{QWS}
\end{center}
\end{figure}

Having established the temporal stability of the interface and Pb layers we now turn to the electronic structure of the quantum well states.
Figure \ref{QWS}(a-c) shows the in-plane dispersions of QWS for Pb film thicknesses of 10, 17 and 19 ML grown on the Bi interface measured along the high symmetry direction $\overline{\text{M}}$-$\overline{\Gamma}$-$\overline{\text{M}}$ of the (1$\times$1) SBZ.
In contrast to Pb films grown on the Pb-$\sqrt{3}$ interface, as shown in Fig. \ref{QWS}(d), we do not observe states with an effective mass larger than 4 $m_{e}$. The effective mass was determined quantitatively by taking EDCs at momentum steps of 0.035 \AA$^{-1}$ for a total range of $\overline{\Gamma}\pm 0.3$\AA$^{-1}$ and fitting them with \emph{Voigt} functions after subtracting a linear background contribution.
From these fits one obtains the QWS energy versus in-plane momentum, which are finally fitted assuming a parabolic dispersion to determine $m^{\star}$.
 
It should be pointed out that all measured QWS bands presented in this work are derived from $6p$ orbitals, because the binding energies lie above the Pb $sp$ symmetry band gap of the $\overline{\Gamma}$-$\overline{\text{L}}$ direction, which expands from 4 to 8 eV binding energy \cite{Horn:30PRB}. Below this gap the QWS are $6sp_{z}$ derived. The electronic states described by a parabolic dispersion around the SBZC have $6p_{z}$ orbital character and the downward dispersing bands at $k_{||}>0.4$ \AA$^{-1}$ have $6p_{xy}$ character \cite{Dil:2007}.

It is remarkable that, independent on the Pb layer thickness, all measured QWS exhibit an enhanced in-plane dispersion compared to Pb/Pb/Si(111)). The QWS displayed in Fig. \ref{QWS}(a) which is found in a 10 ML thick film of Pb has an effective mass of 2.75 $m_{e}$ and the QWS which arises in 19 ML Pb has an effective mass of 2.09 $m_{e}$. This trend indicates that the coupling to the substrate is reduced, even for the low coverage, where largest effective masses were reported for Pb/Pb/Si \cite{Dil:2006}. Figure \ref{QWS}(f) shows the development of the effective mass of $6p_{z}$ derived states with increasing Pb layer thickness for the Pb and the investigated Bi interface.
In Pb/Pb/Si(111) the effective mass decreases with thickness from $m^{\star} = 10$ $m_{e}$ at low coverage towards $m^{\star}$ = 4 $m_{e}$ at 22 ML. With increasing film thickness the effective mass shows a trend towards the bulk limit (N$\rightarrow \infty$, grey dashed line) of $m^{\star}$~=~1.14 $m_{e}$ \cite{Zhang:1989}. Hence for low coverage the influence of the substrate and interface is most evident and with increasing thickness this influence becomes smaller.

\begin{figure}[!h]
\begin{center}
\includegraphics[width=0.48\textwidth]{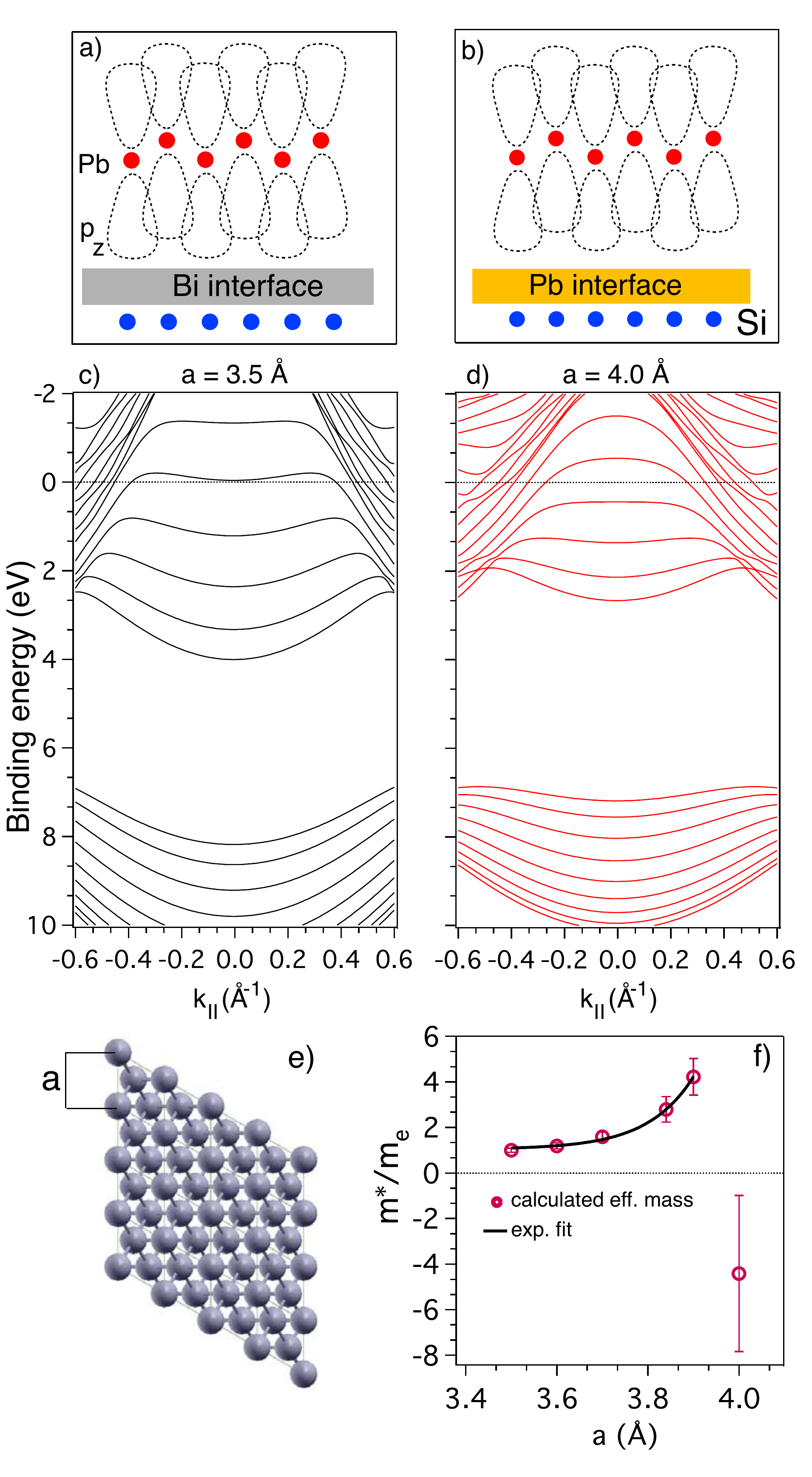}
\caption{(Color online) (a) and (b) Simplified picture of $p_{z}$ orbital overlap at both interfaces. (c) Calculated band dispersion along $\overline{\text{M}}$-$\overline{\Gamma}$-$\overline{\text{M}}$ with in-plane lattice constant $a$ = 3.5 \AA, and (d) $a$~=~4.0~\AA. (e) top view of the Pb(111) surface and (f) effective masses for different in-plane lattice constants obtained from DFT including an exponential fit as described in the text.}
\label{DFT}
\end{center}
\end{figure}

The substrate may force the atoms in the Pb overlayer into an in-plane lattice constant which coincides more with the lattice structure of the substrate. Thus the lattice constant of the overlayer appears for low coverage to be closer to that of Si than to that of Pb. As a consequence, and because the Pb films grow as close packed layers in $fcc$ stacking along the [111] direction, an increase of the interatomic distance between nearest neighbor atoms will decrease the overlap of the $6p_{z}$ orbitals, as drawn schematically in Figs. \ref{DFT}(a) and (b). This in turn will lead to more localized states with flat dispersion. The  influence of the substrate becomes less for thicker films, therefore the lattice constant decreases with thickness. 
Our interpretation that the degree of orbital overlap influences the band dispersion of Pb QWS is also supported by DFT calculations, which were performed using the \emph{ab initio} \emph{Wien2K} simulation package \cite{Schwarz:DFT}. The generalized gradient approximation of Perdew \emph{et al.} \cite{Perdew:1996} was employed to a repeated slab geometry (8 ML Pb, 15 \AA\ vacuum) with a Muffin-tin radius set to 2.5 bohr, plane wave cut-off of 7.84 Ry and integration over the Brillouin zone with (16$\times$16$\times$1) k-points. Spin-orbit interaction was included in a second self-consistent cylce using pertubation theory with the scalar-relativistic orbitals as basis set. Figs \ref{DFT} (c) and (d) show two band dispersions along $\overline{\text{M}}$-$\overline{\Gamma}$-$\overline{\text{M}}$ for the in-plane lattice constants $a$ = 3.5 and 4.0 \AA\ as defined in Fig. \ref{DFT}(e). 

The increase of the in-plane lattice constant by 0.5~\AA\ affects the entire band dispersion of the system. Both the 6$p_{z}$ derived bands and the 6$sp_{z}$ derived bands below the symmetry band gap of Pb become flatter throughout the whole energy range. However, the change of the dispersion of the 6$p_{z}$ states is most significant, which can be assigned to an enhanced influence of the hybridization with the downward dispersing 6$p_{xy}$ states. With the increasing in-plane lattice constant, the band width of these bands is reduced and in the energy range of interest they thus move down in energy and become flatter. The hybridization with the $p_{z}$ bands, and the associated change from electron-like to hole-like dispersion in these bands, therefore starts at lower $|\textbf{k}_{\parallel}|$.

In the following analysis we will focus on the first band below the Fermi level, as this is the state we measured with ARPES. The effective mass of this band changes from initial 1.02 $m_{e}$ to a hole-like dispersion with $m^{\star}$ of -4.10 $m_{e}$ when the lattice constant is increased by 0.5~\AA . Fig. \ref{DFT} (f) shows calculated effective masses for lattice parameters between 3.5 and 4.0 \AA, which are fitted for positive values with an exponential function according to: $m^{\star}(a [$\AA$]) = 1.038+0.0585\exp(\frac{a-3.5}{0.1})$ in units of the free electron mass. Note that the influence of the lattice parameter on the effective mass is weak up to $a = 3.7$~\AA\ and becomes stronger for higher values. This anomalous dependency of the effective mass on the in-plane lattice constant is thus not only a result of the reduced orbital overlap but also strongly enhanced by the hybridization of the 6p$_{z}$ and 6p$_{xy}$ derived bands.

\begin{figure}[h]
\begin{center}
\includegraphics[width=0.45\textwidth]{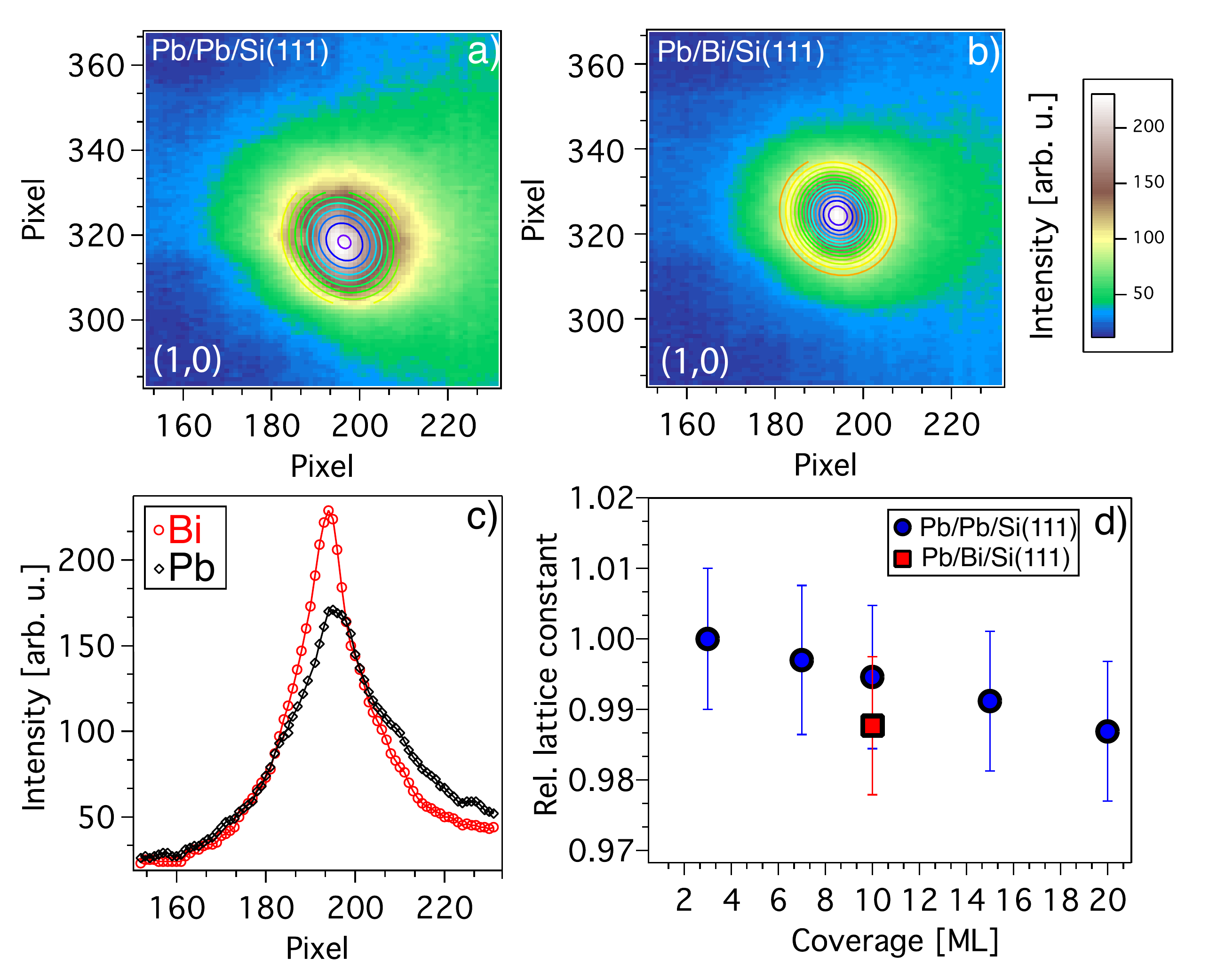}
\caption{(Color online) (a) and (b) (1,0) LEED spot of the Pb/Pb/Si(111) and Pb/Bi/Si(111) surface. (c) Horizontal cut through the maximum of intensity of the (1,0) spot for Bi (red open circles) and Pb (black solid circles). (d) Relative lattice constant vs. coverage measured with LEED; blue circles represent Pb/Pb/Si(111) and the red square Pb/Bi/Si(111).} 
\label{ML}
\end{center}
\end{figure}

\begin{figure}[!h]
\begin{center}
\includegraphics[width=0.5\textwidth]{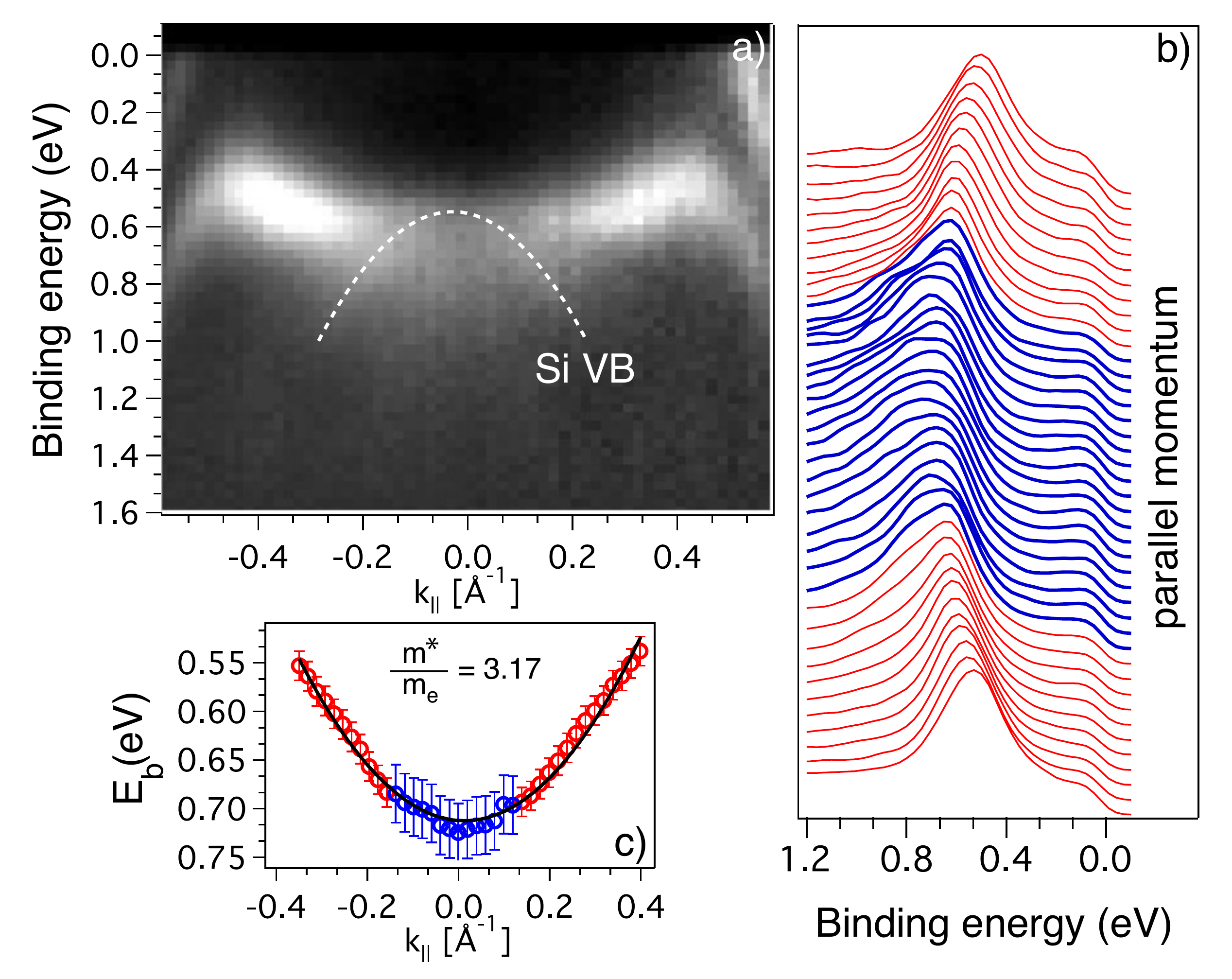}
\caption{(Color online) (a) Band dispersion of 12 ML Pb deposited on top of Bi/Si(111). (b) Normalized energy distribution curves for in-plane momenta k$_{\parallel}=0\pm 0.4$\AA$^{-1}$. Blue spectra represent quantum resonances and red (grey) spectra fully confined QWS. (c) Parabolic fit including the quantum resonances.}
\label{Hybridization}
\end{center}
\end{figure}

Having confirmed with DFT the influence of the in-plane lattice constant on the band dispersion we now turn to the measurements of the surface in-plane lattice constants using LEED, which is sensitive to the surface atomic structure and thus allows the determination of changes in the in-plane lattice constant of the topmost layers. We have recorded the $(1\times 1)$ LEED patterns of the Pb surface for different coverages of Pb on Pb-$\sqrt{3}$/Si(111) and for a 10 ML thick Pb film on Bi-$\sqrt{3}$/Si(111). A comparison of the (1,0) spots for the 10 ML thick film grown on the different interfaces indicates that the spots are sharper on the Bi interface than on the Pb interface as shown in the raw data of Figs. \ref{ML}(a) and (b) and the profiles in Fig. \ref{ML}(c). Sharp LEED spots are either found for perfectly commensurate systems or when the interaction between the substrate and the overlayer is minimal \cite{Dil:2007}.
For a more quantitative determination of the in-plane lattice constant the LEED data were analysed as follows: first, a $2D$ gaussian fit as shown in Figs. \ref{ML}(a) and (b) determines the position of each spot and the corresponding (x,y) coordinates on the screen. Second, from these coordinates the distances between all three $\overline{\text{M}}$-$\overline{\Gamma}$-$\overline{\text{M}}$ directions were determined, which are finally averaged. Although our LEED set-up is sensitive to relative changes of the lattice constant it is not calibrated for the precise determination of absolute values. In the following we therefore use values relative to the atomic spacing found for the lowest thickness of Pb on Pb-$\sqrt{3}$/Si(111). The findings are summarized in Fig. \ref{ML}(d) as a function of layer thickness. The thinnest Pb layer of 3 ML on the Pb interface has an almost 2\% larger lattice constant than the 20 ML Pb film on the same substrate. Using the relation between lattice constant and effective mass obtained from DFT we can calculate the corresponding lattice constant for the 20 ML thick Pb film ($m^{\star}$ = 5 $m_{e}$) which is 3.921 \AA\ (Fig. \ref{DFT}(f)), and estimate the effective mass for this lattice constant expanded by 2\% (3.999 \AA). This gives us $m^{\star}$ = 9.689 $m_{e}$  which is in very good agreement with what is observed for low coverages. The lattice constant of the 10 ML Pb film grown on the Pb-$\sqrt{3}$/Si(111) differs by about 1\% from the Pb film grown on the Bi interface. The comparison of these effective masses with our fit function leads to a lattice constant difference of almost 3\%, which is consistent within our error margins.\\
Upton \emph{et al.} have attempted to explain the unusual high effective mass in Pb/Pb/Si(111) in terms of an avoided crossing of $p_{z}$ derived states and valence states of Si(111) and the rapid oscillations of the phase shift in this energy region  \cite{Upton:2005}. QWS with energies close to the confinement edge $E_{0}$ of the substrate will feel a repulsive force which tends to lower the binding energy and so to prevent a band crossing. 
Figure \ref{Hybridization}(a) shows the band dispersion of a 12 ML Pb film grown on the Bi-$\sqrt{3}$ interface with a binding energy of $E_{B}$~=~690 meV at the zone center and with an effective mass of $m^{\star}$ = 3.17 $m_{e}$.

The parabolic band dispersion of the Pb film intersects with the VB of Si around an in-plane momentum of $k_{\parallel} = (0\pm 0.2)$ \AA$^{-1}$ and becomes less pronounced in this region. Electronic states further away from the VB are not affected. The interaction of film states with the VB of the substrate was also reported for Al/Si(111) \cite{Aballe:2001} and Ag/Ge(111) \cite{Carbone:80PRB} and explained in terms of hybridization of the film and valence states of the substrate. States within the specific momentum range around $\overline{\Gamma}$ are quantum resonances, because they are not confined within the absolute band gap of Si(111). The hybridization leads to a lower intensity and to a larger peak width because the reflectivity of the interface is no longer equal to one. Figure \ref{Hybridization}(b) shows EDCs taken from $k_{\parallel} = -0.4 ... +0.4 $ \AA$^{-1}$ and normalized to the maximum of intensity. Red (light grey) EDCs represent truly confined states (QWS) with narrow line widths and blue (dark grey) EDCs quantum resonances. Also for the Bi reconstructed interface a clear hybridization between the QWS and the Si VB is thus observed. However, the effective mass is still significantly lower compared to QWS on Pb-$\sqrt{3}$/Si(111). Therefore the high effective mass for Pb/Pb/Si(111) can not be explained by the influence of the Si VB alone. The increase of the effective mass actually occurs in the energy and momentum space which is further away from the Si valence band and closer to the 6$p_{xy}$ derived states. Thus the unusual effective mass in Pb/Pb/Si(111) arises from the structural and electronic properties of the Pb overlayer and can be reduced by replacing the Pb with a Bi interface.  
\section*{SUMMARY}
The influence of the Bi interface on the in-plane dispersion of $6p_{z}$ derived QWS has been studied with ARPES and LEED experiments and DFT calculations. In contrast to Pb/Pb/Si(111) we find: $i)$ no effective masses larger than 4 $m_{e}$ and $ii)$ no significant dependence of the effective mass on the layer thickness. Compared to Pb/Pb/Si(111) the dispersion of $6p_{z}$ derived states in Pb/Bi/Si(111) is enhanced. With LEED we could confirm that in Pb/Pb/Si(111) the in-plane lattice constant of the ultra-thin film decreases with coverage in a similar manner as the effective mass. This is fully consistent with our DFT calculations and the corresponding picture of orbital overlap; the reduced overlap of the 6$p_{z}$ orbitals reduces the dispersion, simultaneously the band width of the 6$p_{xy}$ derived states is reduced. Together these effects alter the hybridization between the 6$p_z$ and 6$p_{xy}$ derived bands and cause the change from electron-like to hole-like dispersion to shift to lower $|\textbf{k}_{\parallel}|$.  Furthermore the in-plane lattice constant of a Pb film on the Bi reconstructed interface is smaller compared to a film of similar coverage on the Pb reconstructed interface, which is also reflected in the effective mass. These findings let us conclude that the Bi interface acts as a decoupling layer between the Pb and Si, in the sense that the influence of the substrate on lateral atomic positions of the Pb overlayer is reduced. 

Assistance with the DFT calculations by P. Blaha and R. Laskowski is gratefully acknowledged. 
We thank C. Hess, F. Dubi, and M. Kl\"ockner for technical support. This work is supported by the Swiss National Foundation.

\end{document}